\documentclass{article}

\usepackage[nonatbib,final]{neurips_2024_ml4ps}
\usepackage{color} 
\usepackage{graphicx}
\usepackage[numbers]{natbib}
\newcommand{\etal}{\textit{et~al}.}

\newcommand{\eg}{\textit{e}.\textit{g}.}

\newcommand{\name}[1]{#1}
\newcommand{\parspace}{\vspace{-0.25em}}
\usepackage[utf8]{inputenc} % allow utf-8 input
\usepackage[T1]{fontenc}    % use 8-bit T1 fonts
\usepackage{hyperref}       % hyperlinks
\usepackage{url}            % simple URL typesetting
\usepackage{booktabs}       % professional-quality tables
\usepackage{amsfonts}       % blackboard math symbols
\usepackage{nicefrac}       % compact symbols for 1/2, etc.
\usepackage{microtype}      % microtypography
\usepackage{xcolor}         % colors
\newcommand{\papertitle}{Towards Agentic AI on Particle Accelerators}
\newcommand{\paperabstract}{As particle accelerators grow in complexity, traditional control methods face increasing challenges in achieving optimal performance. This paper envisions a paradigm shift: a decentralized multi-agent framework for accelerator control, powered by Large Language Models (LLMs) and distributed among autonomous agents. We present a proposition of a self-improving decentralized system where intelligent agents handle high-level tasks and communication and each agent is specialized to control individual accelerator components.

This approach raises some questions: What are the future applications of AI in particle accelerators? How can we implement an autonomous complex system such as a particle accelerator where agents gradually improve through experience and human feedback? What are the implications of integrating a human-in-the-loop component for labeling operational data and providing expert guidance? We show three examples, where we demonstrate the viability of such architecture.}

\title{\papertitle{}}

\author{Antonin Sulc \\
  Helmholtz Zentrum Berlin\\
  Berlin, Germany\\
  \texttt{antonin.sulc@helmholtz-berlin.de} \\
  \And
  Thorsten Hellert \\
  LBNL,\\
  Berkeley, USA\\
  \texttt{thellert@lbl.gov} \\
  \And
  Raimund Kammering \\
  DESY,\\
  Hamburg, Germany\\
  \texttt{raimund.kammering@desy.de} \\
  \And
  Hayden Hoschouer\\
  FNAL, \\
  Batavia, IL, USA\\
  \texttt{haydenh@fnal.gov}
  \And
  Jason St. John\\
  FNAL, \\
  Batavia, IL, USA\\
  \texttt{stjohn@fnal.gov}
}

\begin{document}

\maketitle

\begin{abstract}
\paperabstract{}
\end{abstract}

\section{Introduction}
\parspace
Particle accelerator operation involves a complex interplay of interrelated systems, each requiring precise control and optimization. Traditionally, these systems have been managed through a combination of human expertise and highly specialized algorithms with recent developments including machine learning techniques~\cite{edelen2024machine,edelen2018opportunities} where some are successfully deployed in operation like \eg{}~\cite{eichler2021first,PhysRevAccelBeams.27.074602,zhang2022badger}.

Most of these algorithms are designed to operate in the narrow domain of accelerator functionality. The systems have noise, tight operational limits, and deterministic interplay through the control system to ensure human and equipment safety, preventing scenarios that could lead to hazardous conditions or equipment damage.
While effective, this approach can sometimes struggle to achieve optimal overall performance due to the challenges of integrating these disparate systems. Even when a sufficiently comprehensive model of the facility is constructed, it can be susceptible and prone to problems as the accelerator transitions through different states or experiences drifts and thus requires human intervention for setup and ongoing maintenance.

\begin{figure}
	\centering
	\includegraphics[width=1.0\linewidth]{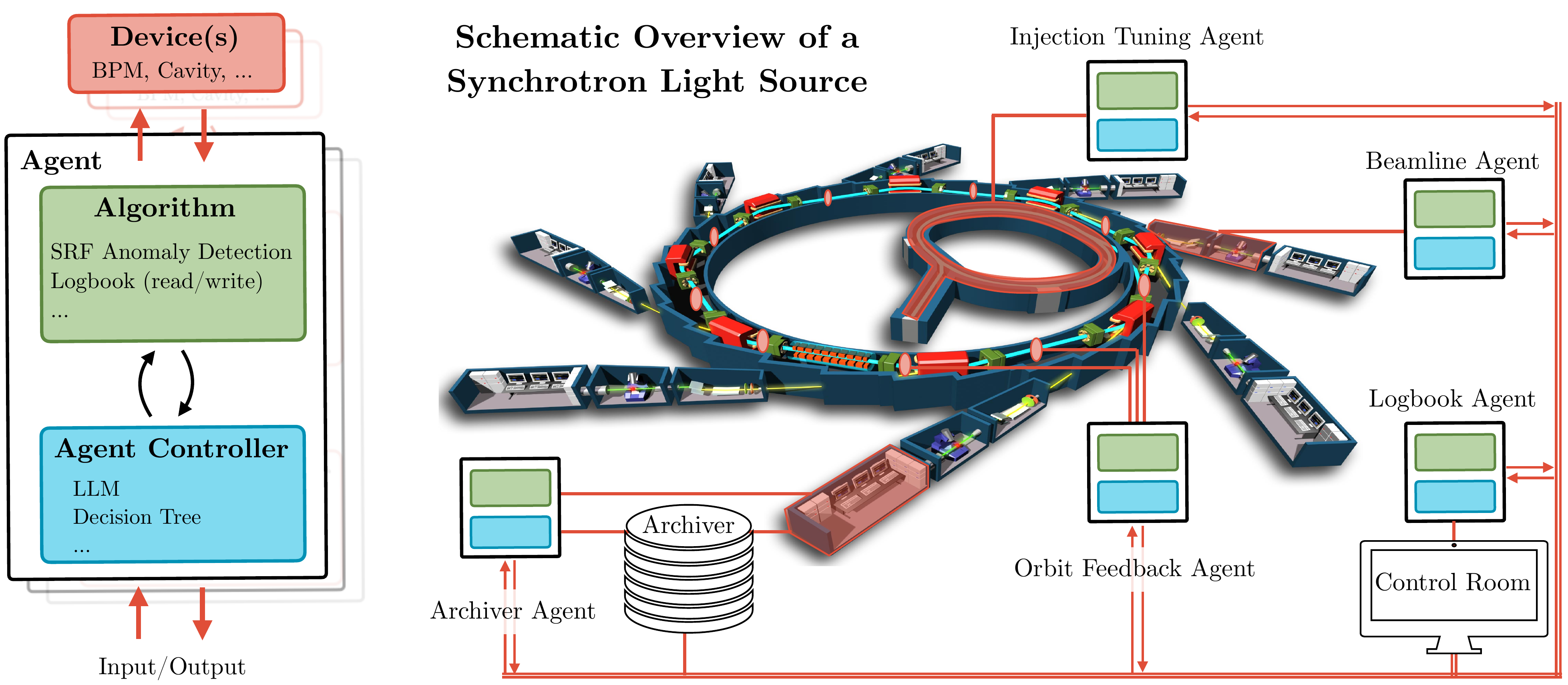}
	\caption{Schematic overview of a decentralized agent-based control architecture. The left diagram illustrates the modular structure where each agent, consisting of devices, algorithms, and agent controllers, manages specific subsystems.
	The right figure shows these agents interacting with both the physical components of the accelerator and the control room, enabling prompting current state, monitoring, decision-making, and adaptive responses to complex operational scenarios.}
	\label{fig:architecture}
\end{figure}

This paper proposes a paradigm shift in accelerator control systems, envisioning a future where self-improving agents control separated sub-components using algorithms that inform operators of significant events.
In this framework, we show how autonomous agents controlled by LLMs serve not only in the creation and retrieval of accelerator operation documentation but also as high-level coordinators, handling complex tasks such as interpreting the machine state from analyzing control system channels or inter-agent communication with a degree of autonomy in decision-making and communication.
Meanwhile, dedicated agents controlled by LLMs execute specific, well-established calculations or operations within their domains of expertise.
This decentralized approach, illustrated in our conceptual diagram Fig.~\ref{fig:architecture}, showcases a potential future for the operation of particle accelerators with the flexibility to deploy more advanced agents.

By exploring this forward-looking perspective, we aim to stimulate discussion on the future of particle accelerator operations and the broader implications of decentralized AI architectures in scientific instrumentation. The proposed framework offers potential solutions to the challenges of integrating disparate systems and managing complex state transitions, while also providing a flexible architecture that can adapt to evolving requirements and technological advancements.

\parspace
\section{Related Work}
\parspace
Machine learning techniques have increasingly been applied to various aspects of accelerator physics in recent years. Edelen~\etal{}~\cite{edelen2024machine,edelen2018opportunities} provide a comprehensive review of machine learning applications in particle accelerators, covering areas such as beam diagnostics, control systems, and performance optimization.

\parspace
\subsection{Agentic AI and LLMs in Complex Reasoning Tasks}
\parspace
There is a rise of agentic AI that has demonstrated remarkable capabilities in complex reasoning tasks~\cite{carrasco2024space}.
%\cite{wang2024survey, xi2023rise}
According to~\cite{xi2023rise}, an effective agent in an agentic system should possess autonomy in independent operation, reactivity in environmental response, and proactiveness in pursuing its objectives, enabling it to function intelligently and adaptively in complex environments.
We will highlight the most relevant works that align with our problem.

Yao~\etal{}~\cite{yao2022react} introduced the \name{ReAct} framework, which showed how LLMs can effectively combine reasoning and acting in multi-step tasks and is widely used to operate in complex multi-step reasoning.
The use of AI agents to effectively engage with their environment and complete a wide array of tasks has been gaining increasing popularity.
Park \etal{}~\cite{park2023generative} introduce \emph{generative agents} that use LLMs to simulate human behavior in interactive environments, showcasing how LLMs can be utilized to model intricate multi-agent interactions and environmental complexities.
Wang~\etal{}~\cite{wang2023voyager} uses an LLM-powered agent to autonomously explore, acquire skills, and make discoveries in unfamiliar environments using components like an automatic curriculum and skill library.
\name{AgentVerse}~\cite{chen2023agentverse} shows a framework for collaborative groups of LLM-powered expert agents, showing improved performance over single agents in various tasks.
This multi-agent approach leverages LLMs for high-level tasks while enabling specialized components to work together autonomously on complex problems.
\name{AutoGen}~\cite{wu2023autogen} framework allows building LLM applications using multiple conversable agents. They show that autonomous, communicating agents can control subcomponents of a system.
Particle accelerator control often utilizes operation sequencing~\cite{bolshakov2003synoptic,frohlich2022taskomat}, there \name{LLM+P}~\cite{liu2023llm+} can help by translating natural language into formal planning languages, potentially bridging the gap between planning, natural language, and accelerator controls.
To deal with environment changes, Shinn~\etal{}~\cite{shinn2024reflexion} proposes \name{Reflexion}, a framework that reinforces language agents through linguistic feedback rather than traditional weight updates.
\name{Reflexion} agents verbally reflect on task feedback and maintain reflective text in an episodic memory buffer to improve decision-making in subsequent trials.
To impose the factuality, \cite{du2023improving} proposes \textit{multiagent debate} where multiple LLMs debate to improve reasoning and factual accuracy of results which is essential for reliable operations.
LLMs have gained prominence as coding assistants~\cite{jiang2024survey} demonstrating the capability to function as coders~\cite{qian2024chatdev}, problem solvers~\cite{FunSearch2023} or even data scientists~\cite{lai2023ds}.
Automated Design of Agentic Systems~\cite{hu2024automated} aims to automatically create AI agent designs by having a meta-agent iteratively program new agents in code.

\parspace
\subsection{LLMs as Assistants in Operations}    
\parspace
Carrasco~\etal{}~\cite{carrasco2024space} explores fine-tuned LLMs for autonomous spacecraft control in simulations, showing their efficacy in handling language-based inputs and outputs. It shows the potential of using LLMs for complex tasks like accelerator control.

In accelerators, Mayet's~\cite{mayet2024gaia} \name{GAIA} system uses LLMs with the ReAct framework~\cite{yao2022react} to assist in operations by integrating multiple expert tools \eg{} knowledge retrieval, machine control, and Python script generation for autonomous and semi-autonomous management of complex accelerator environments.

In~\cite{sulc2024towards} they show joint efforts across particle accelerator facilities to utilize Retrieval Augmented Generation (RAG) models and other AI techniques for enhancing eLogs usability and automation, aiming to unlock operational insights and improve data accessibility.

\parspace
\section{Features of the System}
\parspace
Four key aspects of agents can significantly enhance operational effectiveness.
First, operating particle accelerators is complex and requires extensive human expertise.
We suggest that agents can gradually improve through experience~\cite{shinn2024reflexion} where the system incorporates continuous learning from operational data and \textbf{learns} based on outcomes or from "\emph{human-in-the-loop}". It can be an important step forward since most of these particle accelerators operate in already high reliabilities above 90\%.

Second, the potential to uncover \textbf{causal relationships} in accelerator operations by adding \name{Reasoning Agents}~\cite{han2024causal}.
LLM-powered agents can provide a human-readable interface to complex machine operations, enabling faster exploration of causal connections via reasoning about the information available (or gradually reasoning about it via ReAct~\cite{yao2022react}).
By integrating agents designed to reveal casual relationships, we aim to enhance system interpretability and streamline diagnostics. This approach could reduce the learning curve for new operators and offer valuable insights to experienced physicists, leading to more efficient accelerator management.

The third key component is agent \textbf{autonomy} implementable via rules, decision trees, or language models. LLMs are currently preferred due to their natural language interface and intelligent decision-making.
LLM's high computational needs may cause delays, limiting use in real-time control. The choice of autonomy implementation thus requires balancing LLMs' flexibility against performance needs in time-sensitive applications.

Lastly, other interesting examples of agents can be: planning agent~\cite{liu2023llm+} to execute and run complex plans with standard tools, such as to execute tasks at \eg{} European XFEL defined in \name{Taskomat}~\cite{frohlich2022taskomat} or \cite{bolshakov2003synoptic} at Fermi, coding agents~\cite{qian2024chatdev}, and data scientist agents~\cite{lai2023ds} to summarize.

\parspace
\section{Examples}
\parspace
In this section, we present three examples, where the aforementioned features can be directly applicable by integrating them into the control systems like~\cite{KNOTT1994486,hensler1996doocs,patrick2006fermilab} and enhancing operational efficiency and stability: the Advanced Light Source (ALS) orbit feedback system, the European XFEL longitudinal feedback manager and Fermi coding assistant. 

\parspace
\subsection{ALS Example: Orbit Feedback}
\parspace
Maintaining precise control over the beam orbit through orbit feedback systems is crucial for the stable operation of a storage ring. However, this task can be complicated in practice by machine drifts caused by environmental factors or by maintenance activities that alter the operational characteristics of specific sections of the ring. Under the current operational paradigm, when the orbit feedback system fails to converge during user operation, a physicist must intervene. This process typically involves a detailed analysis of the situation, comparing current conditions with previous runs, and relying heavily on experience to make a judgment call on how to adjust the system for optimal performance.

\begin{figure}[htb]
	\centering
	\includegraphics[width=1\linewidth]{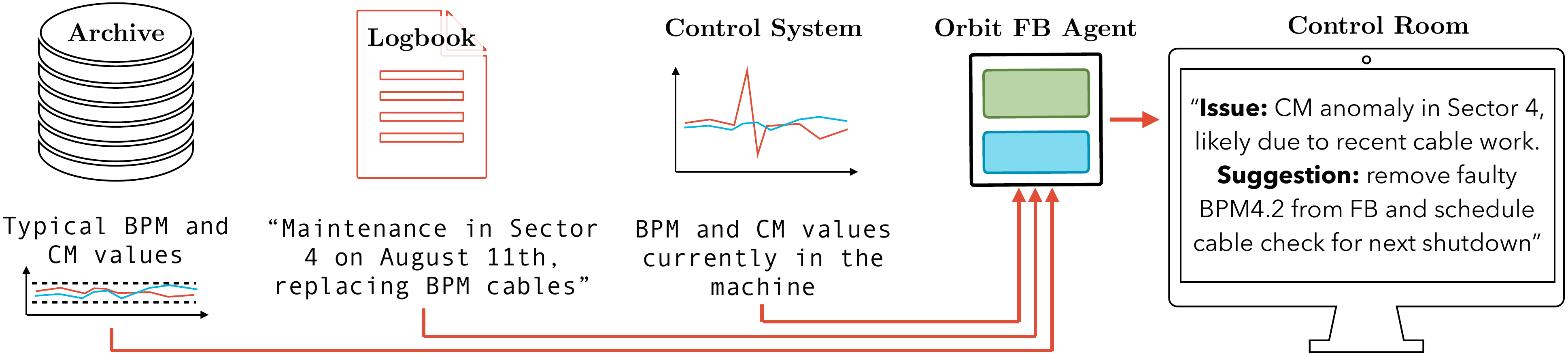}
	\caption{Diagram of the orbit feedback agent. The \name{Orbit FB Agent} detects abnormal behavior, consults the \name{Logbook Agent} for relevant events, and suggests an action for the control room to review.}
	\label{fig:example_ALS_FB}
\end{figure}

As illustrated in Fig.~\ref{fig:example_ALS_FB}, in our envisioned framework, this diagnostic role would be managed by a specialized feedback agent. Upon detecting abnormal behavior, the agent would identify the underperforming area and consult the logbook agent for recent events that could explain the issue. For instance, if maintenance work had occurred in the affected sector, the logbook agent would provide this information. The feedback agent would then draft a report outlining the problem, its likely root cause, and a suggested course of action, which would be sent to the control room for review.

\parspace
\subsection{European XFEL Example: Longitudinal Feedback Manager} \label{subsec:XFEL_feedback}
\parspace

\begin{figure}[htb]
	\centering
	\includegraphics[width=1\linewidth]{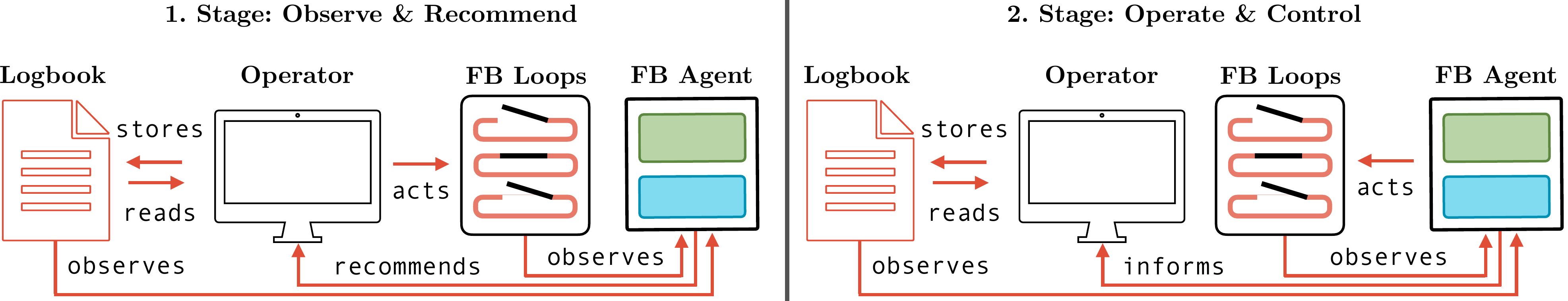}
	\caption{Diagram of the Longitudinal feedback manager. Left: The operator sets feedback loops and the FB Manager Agent and Logbook Agent observe and recommend actions to the operator. Right: FB Manager Agent takes over the operation and informs the operator.}
	\label{fig:example_XFEL}
\end{figure}

The particle beam in the linear accelerator, like the European XFEL, is stabilized within well-defined, narrow parameter spaces through the use of various feedback loops.
To effectively orchestrate the complex interplay of these feedback loops, expert systems are often employed~\cite{kammering2013feedbacks,dinter2018longitudinal}, which encapsulate the logic of this interaction in software.
A \name{Feedback Agent} can be trained to learn the possible and desired states, illustrated in Fig. \ref{fig:example_XFEL}.
In the next step, such an agent would initially be able to assist the operator as a kind of recommender system and, potentially, could take over the entire operation of all longitudinal feedback loops in a later step.

Examples of additional agents can be: \name{Orbit Feedback Agent} for controlling transversal beam parameters; \name{Undulator Agents} for optimizing photon beam production; \name{Beamline Agents} overseeing experimental stations; and \name{Feedback Coordination Agent} acting as a central coordinator to orchestrate overall feedback operation.
Each agent functions as a recommender system, assisting human operators in decision-making. As confidence in their performance grows, they could potentially take over the entire operation of individual feedback, \eg{} through \cite{shinn2024reflexion}.
Crucially, these agents maintain constant communication with each other. Before implementing any changes, each agent informs the others of its intended actions. This allows for a collective evaluation of the proposed changes, ensuring that they don't lead to suboptimal or dangerous configurations.

\subsection{Fermilab Example: ACL Coding Assistant}
Accelerator Command Language (\name{ACL}), developed at Fermilab, is a scripting language designed for non-programmers, such as engineers and accelerator operators, to control complex accelerator systems.
Despite its specialized nature, AI agents can assist with ACL, even without specific training on the language itself.

A \name{Code Retrieval Agent} utilizes a preexisting semantic search API to query ACL's extensive documentation, retrieving relevant code snippets based on detailed descriptions. After gathering these results, the \name{Retrieval Agent} reviews the information and selects the most relevant snippets to fulfill the user’s request.
These snippets are then passed to a \name{Code Generation Agent}, which uses the retrieved information to infer the correct ACL code structure and generate context-aware script suggestions using \eg{}~\cite{guo2023empowering}.

This two-step process simplifies working with ACL and provides intelligent coding assistance, helping users efficiently generate scripts for a proprietary language specific to Fermilab.

However, one potential challenge is that the initially generated code may not always be correct. To address this, combining Reflexion~\cite{shinn2024reflexion} for verbal reinforcement learning and embeddings for long-term memory such as~\cite{zhang2024codeagent} could help guide the model toward generating fully functional code. This approach would enable the model to iteratively refine its outputs based on feedback and past experiences, ensuring more accurate results over time.

\parspace
\section{Safety Issues and Hallucinations}
\parspace
The tendency of LLMs to hallucinate creates substantial risks for critical system applications. To mitigate potential issues arising from hallucinations and uncertainty, several approaches can be implemented.
These include constraining output using defined grammars or regular expressions~\cite{willard2023efficient}, employing multiple agents to cross-check decisions~\cite{besta2024checkembed}, and integrating real-time sensor data feedback loops.

It's important to emphasize that this proposal is for an AI-based, high-level control system that would operate alongside existing control systems. These existing systems have built-in safety measures for protecting equipment and personnel, which would remain independent of the proposed AI control. Our approach is to begin with passive monitoring and operator suggestions before progressing to limited control of non-critical systems.

Furthermore, fallback mechanisms should automatically revert to traditional control systems if anomalies are detected. Continuous validation against established algorithms and expert decisions will be an integral part of the proof of concept. The primary role of the system in this context remains sensing and informing human operators rather than autonomous control, thereby maintaining critical human oversight in accelerator operations.

\parspace
\section{Conclusion}
\parspace
This paper shows a paradigm shift in particle accelerator control through a decentralized multi-agent framework powered by LLMs.
By integrating AI agents for high-level tasks and specialized agents for component management, we address the increasing complexity of modern accelerator systems.

Three examples demonstrate the potential of AI agents in assisting complex accelerator tasks, opening the way for higher autonomy and posing a question of the use case of intelligent agents for assistance with operating particle accelerators.
This approach offers exciting possibilities for improved performance and our practical examples showcase a vision for more intelligent and adaptive accelerator operations.

\clearpage
\bibliographystyle{plain} % We choose the "plain" reference style
\bibliography{references} % Entries are in the refs.bib file

\begin{thebibliography}{10}

\bibitem{besta2024checkembed}
Maciej Besta, Lorenzo Paleari, Ales Kubicek, Piotr Nyczyk, Robert
  Gerstenberger, Patrick Iff, Tomasz Lehmann, Hubert Niewiadomski, and Torsten
  Hoefler.
\newblock {CheckEmbed: Effective Verification of LLM Solutions to Open-Ended
  Tasks}, June 2024.

\bibitem{bolshakov2003synoptic}
Timofei Bolshakov, AD~Petrov, and Sharon Lackey.
\newblock Synoptic display—a client-server system for graphical data
  representation.
\newblock {\em proceedings of the 2003 ICALEPCS, Gyeongju, Korea}, 2003.

\bibitem{carrasco2024space}
Alejandro Carrasco, Victor Rodriguez-Fernandez, and Richard Linares.
\newblock Fine-tuning llms for autonomous spacecraft control: A case study
  using kerbal space program, 2024.

\bibitem{chen2023agentverse}
Weize Chen, Yusheng Su, Jingwei Zuo, Cheng Yang, Chenfei Yuan, Chi-Min Chan,
  Heyang Yu, Yaxi Lu, Yi-Hsin Hung, Chen Qian, et~al.
\newblock Agentverse: Facilitating multi-agent collaboration and exploring
  emergent behaviors.
\newblock In {\em The Twelfth International Conference on Learning
  Representations}, 2023.

\bibitem{KNOTT1994486}
L.~Dalesio, J.~Hill, M.~Kraimer, S.~Lewis, D.~Murray, S.~Hunt, W.~Watson,
  M.~Clausen, and J.~Dalesio.
\newblock The experimental physics and industrial control system architecture:
  past, present, and future.
\newblock {\em Nuclear Instruments and Methods in Physics Research Section A:
  Accelerators, Spectrometers, Detectors and Associated Equipment},
  352:179--184, 1994.

\bibitem{dinter2018longitudinal}
Hannes Dinter.
\newblock {\em Longitudinal diagnostics for beam-based intra bunch-train
  feedback at FLASH and the European XFEL}.
\newblock PhD thesis, Staats-und Universit{\"a}tsbibliothek Hamburg Carl von
  Ossietzky, 2018.

\bibitem{du2023improving}
Yilun Du, Shuang Li, Antonio Torralba, Joshua~B Tenenbaum, and Igor Mordatch.
\newblock Improving factuality and reasoning in language models through
  multiagent debate.
\newblock {\em arXiv preprint arXiv:2305.14325}, 2023.

\bibitem{edelen2024machine}
Auralee Edelen and Xiaobiao Huang.
\newblock Machine learning for design and control of particle accelerators: A
  look backward and forward.
\newblock {\em Annual Review of Nuclear and Particle Science}, 74(1):557--581,
  2024.

\bibitem{edelen2018opportunities}
Auralee Edelen, Christopher Mayes, Daniel Bowring, Daniel Ratner, Andreas
  Adelmann, Rasmus Ischebeck, Jochem Snuverink, Ilya Agapov, Raimund Kammering,
  Jonathan Edelen, et~al.
\newblock Opportunities in machine learning for particle accelerators.
\newblock {\em arXiv preprint arXiv:1811.03172}, 2018.

\bibitem{eichler2021first}
Annika Eichler, Florian Burkart, Jan Kaiser, Willi Kuropka, Oliver Stein, Erik
  Br{\"u}ndermann, Andrea~Santamaria Garcia, and Chenran Xu.
\newblock First steps toward an autonomous accelerator, a common project
  between desy and kit.
\newblock {\em Proc. IPAC’21}, pages 2182--2185, 2021.

\bibitem{frohlich2022taskomat}
L~Fr{\"o}hlich, O~Hensler, U~Jastrow, M~Walla, and J~Wilgen.
\newblock Taskomat \& taskolib: A versatile, programmable sequencer for process
  automation.
\newblock {\em PCaPAC 2022 hosted by ELI Beamlines}, page~94, 2022.

\bibitem{guo2023empowering}
Jing Guo, Nan Li, Jianchuan Qi, Hang Yang, Ruiqiao Li, Yuzhen Feng, Si~Zhang,
  and Ming Xu.
\newblock Empowering working memory for large language model agents.
\newblock {\em arXiv preprint arXiv:2312.17259}, 2023.

\bibitem{han2024causal}
Kairong Han, Kun Kuang, Ziyu Zhao, Junjian Ye, and Fei Wu.
\newblock Causal agent based on large language model.
\newblock {\em arXiv preprint arXiv:2408.06849}, 2024.

\bibitem{PhysRevAccelBeams.27.074602}
Thorsten Hellert, Tynan Ford, Simon~C. Leemann, Hiroshi Nishimura, Marco
  Venturini, and Andrea Pollastro.
\newblock Application of deep learning methods for beam size control during
  user operation at the advanced light source.
\newblock {\em Phys. Rev. Accel. Beams}, 27:074602, Jul 2024.

\bibitem{hensler1996doocs}
O~Hensler and K~Rehlich.
\newblock Doocs: A distributed object oriented control system.
\newblock In {\em Proceedings of XV Workshop on Charged Particle Accelerators,
  Protvino}, 1996.

\bibitem{hu2024automated}
Shengran Hu, Cong Lu, and Jeff Clune.
\newblock Automated design of agentic systems.
\newblock {\em arXiv preprint arXiv:2408.08435}, 2024.

\bibitem{jiang2024survey}
Juyong Jiang, Fan Wang, Jiasi Shen, Sungju Kim, and Sunghun Kim.
\newblock A survey on large language models for code generation.
\newblock {\em arXiv preprint arXiv:2406.00515}, 2024.

\bibitem{kammering2013feedbacks}
Raimund Kammering and Christian Schmidt.
\newblock Feedbacks and automation at the free electron laser in hamburg
  (flash).
\newblock {\em Proc. ICALEPCS’13}, pages 1345--1347, 2013.

\bibitem{lai2023ds}
Yuhang Lai, Chengxi Li, Yiming Wang, Tianyi Zhang, Ruiqi Zhong, Luke
  Zettlemoyer, Wen-tau Yih, Daniel Fried, Sida Wang, and Tao Yu.
\newblock Ds-1000: A natural and reliable benchmark for data science code
  generation.
\newblock In {\em International Conference on Machine Learning}, pages
  18319--18345. PMLR, 2023.

\bibitem{liu2023llm+}
Bo~Liu, Yuqian Jiang, Xiaohan Zhang, Qiang Liu, Shiqi Zhang, Joydeep Biswas,
  and Peter Stone.
\newblock Llm+ p: Empowering large language models with optimal planning
  proficiency.
\newblock {\em arXiv preprint arXiv:2304.11477}, 2023.

\bibitem{mayet2024gaia}
Frank Mayet.
\newblock Gaia: A general ai assistant for intelligent accelerator operations.
\newblock {\em arXiv preprint arXiv:2405.01359}, 2024.

\bibitem{park2023generative}
Joon~Sung Park, Joseph O'Brien, Carrie~Jun Cai, Meredith~Ringel Morris, Percy
  Liang, and Michael~S Bernstein.
\newblock Generative agents: Interactive simulacra of human behavior.
\newblock In {\em Proceedings of the 36th annual acm symposium on user
  interface software and technology}, pages 1--22, 2023.

\bibitem{patrick2006fermilab}
James Patrick.
\newblock The fermilab accelerator control system.
\newblock {\em Proc. ICAP’06}, pages 246--249, 2006.

\bibitem{qian2024chatdev}
Chen Qian, Wei Liu, Hongzhang Liu, Nuo Chen, Yufan Dang, Jiahao Li, Cheng Yang,
  Weize Chen, Yusheng Su, Xin Cong, et~al.
\newblock Chatdev: Communicative agents for software development.
\newblock In {\em Proceedings of the 62nd Annual Meeting of the Association for
  Computational Linguistics (Volume 1: Long Papers)}, pages 15174--15186, 2024.

\bibitem{FunSearch2023}
Bernardino Romera-Paredes, Mohammadamin Barekatain, Alexander Novikov, Matej
  Balog, M.~Pawan Kumar, Emilien Dupont, Francisco J.~R. Ruiz, Jordan
  Ellenberg, Pengming Wang, Omar Fawzi, Pushmeet Kohli, and Alhussein Fawzi.
\newblock Mathematical discoveries from program search with large language
  models.
\newblock {\em Nature}, 2023.

\bibitem{shinn2024reflexion}
Noah Shinn, Federico Cassano, Ashwin Gopinath, Karthik Narasimhan, and Shunyu
  Yao.
\newblock Reflexion: Language agents with verbal reinforcement learning.
\newblock {\em Advances in Neural Information Processing Systems}, 36, 2024.

\bibitem{sulc2024towards}
Antonin Sulc, Alex Bien, Annika Eichler, Daniel Ratner, Florian Rehm, Frank
  Mayet, Gregor Hartmann, Hayden Hoschouer, Henrik Tuennermann, Jan Kaiser,
  et~al.
\newblock Towards unlocking insights from logbooks using ai.
\newblock {\em arXiv preprint arXiv:2406.12881}, 2024.

\bibitem{wang2023voyager}
Guanzhi Wang, Yuqi Xie, Yunfan Jiang, Ajay Mandlekar, Chaowei Xiao, Yuke Zhu,
  Linxi Fan, and Anima Anandkumar.
\newblock Voyager: An open-ended embodied agent with large language models.
\newblock {\em arXiv preprint arXiv:2305.16291}, 2023.

\bibitem{willard2023efficient}
Brandon~T Willard and R{\'e}mi Louf.
\newblock Efficient guided generation for llms.
\newblock {\em arXiv preprint arXiv:2307.09702}, 2023.

\bibitem{wu2023autogen}
Qingyun Wu, Gagan Bansal, Jieyu Zhang, Yiran Wu, Shaokun Zhang, Erkang Zhu,
  Beibin Li, Li~Jiang, Xiaoyun Zhang, and Chi Wang.
\newblock Autogen: Enabling next-gen llm applications via multi-agent
  conversation framework.
\newblock {\em arXiv preprint arXiv:2308.08155}, 2023.

\bibitem{xi2023rise}
Zhiheng Xi, Wenxiang Chen, Xin Guo, Wei He, Yiwen Ding, Boyang Hong, Ming
  Zhang, Junzhe Wang, Senjie Jin, Enyu Zhou, et~al.
\newblock The rise and potential of large language model based agents: A
  survey.
\newblock {\em arXiv preprint arXiv:2309.07864}, 2023.

\bibitem{yao2022react}
Shunyu Yao, Jeffrey Zhao, Dian Yu, Nan Du, Izhak Shafran, Karthik Narasimhan,
  and Yuan Cao.
\newblock React: Synergizing reasoning and acting in language models.
\newblock {\em arXiv preprint arXiv:2210.03629}, 2022.

\bibitem{zhang2024codeagent}
Kechi Zhang, Jia Li, Ge~Li, Xianjie Shi, and Zhi Jin.
\newblock Codeagent: Enhancing code generation with tool-integrated agent
  systems for real-world repo-level coding challenges.
\newblock {\em arXiv preprint arXiv:2401.07339}, 2024.

\bibitem{zhang2022badger}
Z.~Zhang, M.~Böse, A.L. Edelen, J.R. Garrahan, Y.~Hidaka, C.E. Mayes, S.A.
  Miskovich, D.F. Ratner, R.J. Roussel, J.~Shtalenkova, S.~Tomin, and G.M.
  Wang.
\newblock {Badger: The Missing Optimizer in ACR}.
\newblock In {\em Proc. IPAC'22}, number~13 in International Particle
  Accelerator Conference, pages 999--1002. JACoW Publishing, Geneva,
  Switzerland, 07 2022.

\end{thebibliography}

\end{document}